\documentclass{article}

\usepackage{PRIMEarxiv}
\usepackage{makecell}
\usepackage{tabularx}
\usepackage[utf8]{inputenc} 
\usepackage[T1]{fontenc}    
\usepackage[hidelinks,urlcolor=blue]{hyperref}
\usepackage{url}            
\usepackage{booktabs}       
\usepackage{amsfonts}       
\usepackage{nicefrac}       
\usepackage{array}
\usepackage{microtype}      
\usepackage{lipsum}
\usepackage{fancyhdr}       
\usepackage{graphicx}       
\graphicspath{{}}     
\usepackage{textcomp} 
\usepackage{graphicx}
\usepackage{subcaption}
\usepackage{array}
\usepackage[most]{tcolorbox}

\title{Characterising Open Source Co-opetition in Company-hosted Open Source Software Projects: The Cases of PyTorch, TensorFlow, and Transformers}

\author{
  Cailean Osborne\thanks{Corresponding author: Cailean Osborne, cailean.osborne@oii.ox.ac.uk} \\
  University of Oxford \\
  Oxford, UK \\ 
 \And
  Farbod Daneshyan \\
  Peking University \\
  Beijing, China\\
  \And
  Runzhi He \\
  Peking University \\
  Beijing, China\\
 \AND
  Hengzhi Ye \\
  Tsinghua University \\
  Beijing, China\\
  \And
  Yuxia Zhang \\
  Beijing Institute of Technology \\
  Beijing, China\\
  \And
  Minghui Zhou \\
  Peking University \\
  Beijing, China\\
  \AND
}

\begin{document}
\maketitle
\vspace{-1em}

\begin{abstract}
Companies, including market rivals, have long collaborated on the development of open source software (OSS), resulting in a tangle of co-operation and competition known as ``open source co-opetition''. While prior work investigates open source co-opetition in OSS projects that are hosted by vendor-neutral foundations, we have a limited understanding thereof in OSS projects that are hosted and governed by one company. Given their prevalence, it is timely to investigate open source co-opetition in such contexts. Towards this end, we conduct a mixed-methods analysis of three company-hosted OSS projects in the artificial intelligence (AI) industry: Meta's PyTorch (prior to its donation to the Linux Foundation), Google's TensorFlow, and Hugging Face's Transformers. We contribute three key findings. First, while the projects exhibit similar code authorship patterns between host and external companies (\textasciitilde80\%/20\% of commits), collaborations are structured differently (e.g., decentralised vs. hub-and-spoke networks). Second, host and external companies engage in strategic, non-strategic, and contractual collaborations, with varying incentives and collaboration practices. Some of the observed collaborations are specific to the AI industry (e.g., hardware-software optimizations or AI model integrations), while others are typical of the broader software industry (e.g., bug fixing or task outsourcing). Third, single-vendor governance creates a power imbalance that influences open source co-opetition practices and possibilities, from the host company's singular decision-making power (e.g., the risk of license change) to their community involvement strategy (e.g., from over-control to over-delegation). We conclude with recommendations for future research.
\end{abstract}


\section{Introduction}

Companies have participated in the collaborative development of open source software (OSS) since the late 1990s \cite{broca_communs_2021}, capitalising on a myriad of benefits, from lower development costs ~\cite{bonaccorsi_comparing_2006,birkinbine_incorporating_2020} to open standards and interoperability ~\cite{li_systematic_2024,lerner_incentives_2002}. With the increasing prevalence of commercial participation in OSS development, Computer-Supported Cooperative Work (CSCW) researchers have been encouraged to investigate the incentives, roles, and effects of such commercial activity on the norms, practices, and future trajectories of OSS developer communities \cite{germonprez_eight_2018}. Within this field of research, one line of inquiry focuses on why and how companies collaboratively develop OSS \cite{germonprez_open_2013,zhang_how_2020}, including market rivals and companies that are engaged in patent wars against each other \cite{teixeira_cooperation_2016, nguyen_duc_software_2019}. The term ``open source co-opetition'' has been coined to convey this tangle of co-operation and competition between companies \cite{teixeira_understanding_2014}. To date, prior work on open source co-opetition primarily focuses on projects that are hosted by vendor-neutral foundations, such as the OpenStack Foundation\footnote{N.B.: The OpenStack Foundation was renamed the Open Infrastructure Foundation in 2021.} \cite{zhang_companies_2021,teixeira_lessons_2015,teixeira_cooperation_2016}, Linux Foundation (LF)~\cite{germonprez_open_2013}, Apache Software Foundation \cite{linaker_how_2016}, and Eclipse Foundation \cite{wagstrom_vertical_2009}. Yet due to their vendor-neutrality, foundations play a structural role in enabling collaboration between ``unexpected allies'' \cite{omahony_boundary_2008,germonprez_open_2013}, limiting the generalisability of prior work to OSS projects that lack such vendor-neutral governance, such as ones that are initiated, hosted, and governed by one company (henceforth: company-hosted OSS projects).\footnote{N.B.: The literature uses different nomenclature to describe this kind of OSS project: ``company-hosted''~\cite{zhou_inflow_2016}, ``company-managed''~\cite{omahony_emergence_2007}, and ``company-sponsored''~\cite{west_contrasting_2005}. Throughout this study, we use ``company-hosted'' OSS projects to refer to OSS projects that are initiated, hosted, and governed by one company \cite{schaarschmidt_how_2015}.} 

Given the prevalence and impact of company-hosted OSS project across software domains and industries, from web development \cite{thomas_react_2018} to artificial intelligence (AI) \cite{langenkamp_how_2022,osborne_why_2024}, it is timely to address this research gap and advance our understanding of open source co-opetition in projects with different governance models. We address this research gap through a mixed-methods analysis of open source co-opetition in three company-hosted OSS projects in the AI industry: Meta's PyTorch prior to its donation to the LF, Google's TensorFlow, and Hugging Face's (HF) Transformers.  We examine three targeted research questions (RQ), guided by the objective of extending theory on open source co-opetition strategies and practices to the context of company-hosted OSS projects. First, through repository mining and social network analysis (SNA), we investigate \textbf{(RQ1)} patterns of co-opetition, providing a baseline understanding of commonalities and differences between the three cases. Subsequently, through 10 semi-structured interviews, we investigate \textbf{(RQ2)} the types of collaborative relationships that host and external companies pursue and why, as well as \textbf{(RQ3)} what similarities and differences characterise open source co-opetition practices in company-hosted OSS projects compared to foundation-hosted projects, as identified by prior work.

We make three key contributions to the literature on open source co-opetition. First, while code authorship patterns by host and external companies are consistent across the projects over time (e.g., \textasciitilde80\%/20\% of commits respectively),  we observe varying structures of collaboration between companies on project files (e.g., decentralised vs. hub-and-spoke networks). Second, we identify and characterise three distinct relationship types between host and external companies: strategic, non-strategic, and contractual collaborations. Each type differs in the relevance of business strategy, competitive dynamics, and personal incentives for the involved developers. Some of the observed collaborations are specific to the technology and competitive dynamics in AI industry (e.g., hardware-software optimization or AI model integrations), while others are typical of the broader software industry (e.g., bug fixing, code adoption for increased impact, and outsourcing of development). Third, single-vendor governance in company-hosted OSS projects introduces a power imbalance that influences open source co-opetition practices and possibilities, from the host's singular decision-making authority (e.g., the risk of license changes) to their community involvement strategy (e.g., from over-control of development to over-delegation).

The paper has the following structure. First, we discuss prior work on commercial participation in OSS development (Section~\ref{sec:coop-relatedwork}). Next, we present our research design (Section~\ref{sec:coop-researchdesign}). Then, we report the results (Section~\ref{sec:coop-results}). Then, we discuss the key implications of the findings and limitations (Section~\ref{sec:coop-discussion}). Finally, we conclude the paper (Section~\ref{sec:coop-conclusion}).

\section{Related Work} \label{sec:coop-relatedwork}

\subsection{Commercial participation in OSS development}

Companies have collaborated on OSS development since the 1990s \cite{broca_communs_2021}. The last decade, in particular, has seen a ``rapid acceleration of corporate engagement with open source'' \cite{germonprez_eight_2018}, leading to what scholars have called ``the incorporation of the digital commons'' \cite{birkinbine_incorporating_2020} or the emergence of a ``commons of capital'' \cite{calimaq_communs_2018}.  In light of these developments, CSCW researchers have been encouraged to investigate the incentives, roles, and effects of commercial activity \cite{germonprez_eight_2018}. 

Companies participate in OSS development in various ways, which can be broadly categorised into three models: supporting, collaborating, and hosting OSS \cite{zhou_inflow_2016}. In the supporting model, a company assists an independently hosted project. This may include by deploying developers \cite{dahlander_man_2006}, funding maintainers or the project \cite{osborne_public-private_2024}, or joining project steering committees \cite{butler_investigation_2018}, among others. The collaborating model involves multiple organisations sharing control over the project's intellectual property. In the AI industry, this was exemplified by the joint release of the Open Neural Network Exchange (ONNX) by Facebook and Microsoft in 2017, which focused on facilitating  interoperability between multiple deep learning frameworks \cite{candela_facebook_2017}. In the hosting model, a single company exercises full control over a project's governance \cite{yue_igniting_2024} and intellectual property \cite{zhou_inflow_2016}, achieved by employing the maintainers \cite{omahony_emergence_2007} and often requiring contributors to sign contributor license agreements (CLA) \cite{zhou_inflow_2016}, among others. For this reason, O’Mahony \cite{omahony_emergence_2007} refers to such projects as ``company-managed projects'' to underline that they are not  ``initiated and managed by a distributed group of individuals who do not share a common employer''  \cite{omahony_emergence_2007}. Companies spin-out  proprietary software projects into company-hosted OSS projects in order to increase adoption of the software, to benefit from external contributions, or to reduce a competitor’s market share \cite{west_contrasting_2005}.

In company-hosted OSS projects, companies take different approaches to project governance and community involvement in line with their strategic goals \cite{zhou_inflow_2016}. While some companies maintain complete control of their project, which ``resembles proprietary development conducted within a glass house'', others strive to attract contributors, which involves significant investment into community development \cite{west_contrasting_2005}. However, no matter how many resources a company may invest in its project, it is not guaranteed that a company can successfully build and retain a community of contributors in such projects because they are dominated by the company and lack aspects that attract external contributors, such as open community governance or a meritocratic culture \cite{osborne_why_2024,yue_igniting_2024}. What is more, contributors may be suspicious of projects that ``are viewed as transfers to the community to maintain code as opposed to collaborative partnerships'' \cite{west_contrasting_2005}. Prior work also shows that the incentives and types of contributions to company-hosted OSS projects tend to be needs-based, compared to a combination of hobbyist and needs-based contributions by contributors to OSS projects that are led and governed by a community \cite{shah_motivation_2006}.

Companies have diverse incentives for participating in OSS development, which differ from the primarily intrinsic motivations of volunteer contributors \cite{benkler_wealth_2006,von_krogh_carrots_2012}. These incentives encompass both strategic and social benefits. Strategic incentives include reducing development costs \cite{birkinbine_incorporating_2020,crowston_freelibre_2012}, influencing open standards \cite{fink_business_2003,lerner_simple_2002}, and recruiting software engineers \cite{agerfalk_outsourcing_2008,fink_business_2003}. Companies also seek vendor independence \cite{chesbrough_measuring_2023,lerner_simple_2002}, faster time to market \cite{ahlawat_why_2021,chesbrough_measuring_2023}, and enhanced market competitiveness \cite{lindman_beyond_2009,loebbecke_open_2003}. Social incentives involve reciprocating to the OSS ecosystem \cite{feller_understanding_2002,franck_reconciling_2002} and improving corporate reputation as an OSS patron \cite{osterloh_trust_2003,osborne_public-private_2024}. Additionally, companies benefit from OSS adoption through cost savings, faster development speeds, and improved interoperability \cite{chesbrough_measuring_2023}.

These diverse incentives highlight the multifaceted nature of commercial participation in OSS development. However, the collaborative nature of OSS development often leads to scenarios where companies, including market rivals, work together in OSS projects. This phenomenon, known as ``open source co-opetition'' \cite{teixeira_understanding_2014}, represents a unique intersection of co-operation and competition in the OSS context. In the following section, we delve deeper into the concept of open source co-opetition, exploring its definition, manifestations, and the current state of research in this area.

\subsection{Open source co-opetition: Definition and Prior Work}

It is commonplace for companies to collaboratively develop OSS \cite{teixeira_collaboration_2014}. As a result, many OSS communities have evolved ``from networks of individuals to networks of companies'' \cite{agerfalk_outsourcing_2008}. ``Open-source co-opetition'' has been coined to convey the tangle of co-operation and competition between companies in the OSS context \cite{teixeira_understanding_2014}, drawing on the ``co-opetition'' concept from the management science literature \cite{brandenburger_co-opetition_1997,dagnino_coopetition_2009}, which contends that companies, which might be market rivals or even engaged in patent wars, form strategic alliances in areas that far from the customer, such as in research and development (R\&D) in the high-technology sector \cite{stuart_network_1998,dagnino_coopetition_2009}, whilst competing on revenue-generating products and services \cite{bengtsson_coopetition_2000}. These alliances are known as ``access relationships'' that provide access to the resources of other companies \cite{stuart_interorganizational_2000,zineldin_co-opetition_2004}, to facilitate learning \cite{powell_interorganizational_1996,hamel_competition_1991}, and tp enable companies to improve their market position \cite{kogut_joint_1988,stuart_interorganizational_1999} and shape industry standards \cite{gnyawali_co-opetition_2011,brandenburger_rules_2021}. These alliances often create strategic interdependencies between companies \cite{dagnino_coopetition_2009}. 

Scholars have adopted this framework to investigate strategies and practices of inter-company collaboration in OSS development, both at the level of individual developers and companies \cite{nguyen-duc_software_2019,teixeira_cooperation_2016}. At the developer level, prior work finds that open source co-opetition is collaboration-focused; for example, company-affiliated developers report little interest in which companies other contributors work for and typically view contributors from other companies as their peers \cite{nguyen-duc_software_2019}. Furthermore, collaboration is characterised by low affiliation-based homophily, with frequent inter-company collaboration in OSS projects \cite{teixeira_lessons_2015}. Prior work also identifies two mechanisms of competition at the individual developer level. First,  while multiple developers from a company may contribute to an OSS project, there is typically a gatekeeper---which may be one developer or a handful of developers---who coordinates a company’s strategy, files issue reports, submits pull requests (PRs), and manages information flows \cite{nguyen_duc_coopetition_2017,nguyen_duc_software_2019}. The gatekeeper has the authority to decide what information or code the company will share with the project, and therefore acts as a key lever for companies to engage in co-operative and competitive interactions simultaneously \cite{nguyen_duc_coopetition_2017,nguyen-duc_software_2019}. Second, the fork provides developers with the option to deviate at any time to pursue their own strategic goals or if they are not content with the direction of the project \cite{teixeira_collaboration_2014}. The threat of the fork also encourages dominant contributors to appease and retain other contributors \cite{teixeira_collaboration_2014}.

At the company-level, a study on the OpenStack ecosystem found that companies engage in intentional, passive, and isolated collaborations \cite{zhang_how_2020}. Intentional collaborations are collaborations between companies that have a market relationship, such as the supply and consumption of OpenStack software or or service provision \cite{zhang_how_2020}. Passive collaborations occur when companies contribute to the same project without explicit coordination, while isolated collaborations take place when a company contributes to a project alone \cite{zhang_how_2020}. Another study found that companies that share the same revenue model, such as offering complementary software or hardware, collaborated more than those with different revenue models in the OpenStack ecosystem \cite{teixeira_lessons_2015}.

Prior work on open source co-opetition is primarily limited to OSS projects that are hosted by vendor-neutral foundations, including the OpenStack Foundation \cite{zhang_companies_2021,teixeira_lessons_2015,teixeira_cooperation_2016}, LF \cite{germonprez_open_2013}, Apache Software Foundation \cite{linaker_how_2016}, and Eclipse Foundation \cite{wagstrom_vertical_2009}. However, foundation-hosted projects have key characteristics that threaten the generalisability of prior findings to other project hosting and governance models. Through their vendor-neutrality and open governance, foundations operate as ``boundary organisations'' that foster collaboration between diverse contributors, including volunteers and for-profit companies \cite{omahony_boundary_2008}. They are reputed to foster ``communities of competitors,'' where ``market rivals...intentionally coordinate activities for mutual benefit in precise, market-focused, non-differentiating engagements'' \cite{germonprez_open_2013}. We note that foundation-hosted projects do not shield projects from the dominance of a single company \cite{wagstrom_vertical_2009,zhang_companies_2021}. For example, 10\% of companies contribute 80\% of commits and 20\% of companies employ 80\% of the contributors in the OpenStack ecosystem \cite{zhang_companies_2021}. Furthermore, IBM continued to dominate OSS projects hosted in the Eclipse ecosystem long after it established the Eclipse Foundation \cite{wagstrom_vertical_2009}. 
Commercial dominance can have negative consequences for the participation of volunteers \cite{zhou_inflow_2016}, which in part is due to the concern of performing free labour for the dominant company \cite{zhang_companies_2018}.

Beyond OSS projects that are hosted by vendor-neutral foundations, we have a limited understanding of open source co-opetition in OSS projects with different hosting models, such as OSS projects that are initiated, hosted, and governed by one company. To date, only two case studies have investigated open source co-opetition in such scenarios. A network analysis of collaboration in Google's Android simply highlights the dominance of Google developers as well as non-trivial contributions from its market rival Apple \cite{orucevic-alagic_network_2014}, while a study on Apple's WebKit demonstrates the methodological utility of temporal network visualisations for observing evolving collaborations between companies \cite{teixeira_collaboration_2014}. However, these studies fail to investigate open source co-opetition strategies and practices, nor do they consider how different governance approaches taken by host companies influence strategies and practices, from those that maintain control of development \cite{zhou_inflow_2016} to those that adopt a more community-oriented approach \cite{west_contrasting_2005}. Given the prevalence and impact of company-hosted OSS projects across the software industry, it is timely to investigate the strategies and practices of open source co-opetition in such contexts, and ultimately to advance our understanding of the nature and impact of commercial participation in OSS development \cite{germonprez_eight_2018}.

\section{Study Design} \label{sec:coop-researchdesign}

\subsection{Research objectives and research questions} 

Our research objectives are two-fold: first, we seek to test prior theory on open source co-opetition practices in the context of company-hosted OSS projects; and second, to identify and characterise co-opetition practices that are unique to company-hosted OSS projects. Underlying these objectives is our motivating RQ: How do companies co-operate on OSS development when a project is hosted and governed by a single company? We operationalise this motivating RQ by asking three targeted RQs, which we examine through a sequential, mixed-methods analysis of three case studies. First, through repository mining and SNA, we seek to understand \textbf{(RQ1)} typical patterns, if any, of open source co-opetition in company-hosted OSS projects, providing a baseline understanding commonalities and differences between the cases \cite{easterbrook_selecting_2008}. Subsequently, through 10 semi-structured interviews with company-affiliated contributors to the three projects, we seek to identify and characterise \textbf{(RQ2)} different types of collaborative relationships between host and external companies and their motivations, as well as \textbf{(RQ3)} the similarities and differences that characterise open source co-opetition in company-hosted OSS projects compared to foundation-hosted projects. This research design enables the testing of prior theory with mixed-methods findings from multiple cases \cite{easterbrook_selecting_2008,runeson_case_2012}, and enhances the convergence validity of the findings \cite{runeson_guidelines_2008, jick_mixing_1979}. Ethical clearance was obtained from the relevant institutional review board by the primary author prior to the commencement of this research project.

\subsection{Multiple case study research design}

\subsubsection{Case selection}\label{sec:coop-cases-sel}
We employed a four-step strategy to select cases. First, we defined the selection criteria: OSS projects had to be hosted by a company, involve external companies, and undergo active maintenance. Second, we selected the AI industry as the boundaries for case selection due to evidence of commercial investments \cite{ahmed_growing_2023,whittaker_steep_2021} and involvement in the development of OSS \cite{langenkamp_how_2022,osborne_public-private_2024,osborne_why_2024} and open-weight models \cite{white_model_2024,osborne_ai_2024}. 
Third, we prepared a ``starting list'' of company-hosted OSS projects by downloading a database of over 300 OSS projects from the LF AI \& Data Foundation's website~\cite{lfaidata_linux_2022}. We removed data projects, resulting in 184 AI OSS projects. Fourth, we labelled projects according to the type of its hosting organisation (company, foundation, or university) and sorted the projects by the size of their contributor community. Then, we selected the three top-ranked projects in descending order: Google's TensorFlow, Meta's PyTorch, and HF's Transformers. Given Meta's donation of PyTorch to the Linux Foundation in September 2022 \cite{osborne_why_2024}, we limited data collection to this time point to ensure that all projects were company-hosted throughout the analysis period.

The selected cases---PyTorch, TensorFlow, and Transformers---represent different layers of the AI stack. TensorFlow \cite{google_TensorFlow_2023} and PyTorch \cite{PyTorch_PyTorch_2023} are foundational deep learning frameworks used to train machine learning models, while Transformers \cite{huggingface_transformers_2023} provides higher-level APIs for downloading, fine-tuning, and sharing pre-trained models hosted on the HF Hub, a popular platform for hosting and developing AI models and datasets. Given their widespread usage in the AI industry and large contributor communities, they are promising cases for the study of open source co-opetition. In particular, TensorFlow and PyTorch make for interesting comparative cases because Google and Meta have long been in a heated rivalry over industry adoption of their respective deep learning frameworks \cite{oconnor_pytorch_2021}. Furthermore, the inclusion of Transformers enables a comparative analysis of projects that are hosted by industry giants (i.e., Meta and Google) and start-ups (i.e., HF), thus overcoming the limited focus on OSS projects hosted by industry giants in prior work \cite{teixeira_collaboration_2014,orucevic-alagic_network_2014}. We acknowledge the temporal cut-off in September 2022 and focus on popular company-hosted OSS projects with >1,000 contributors as limitations, which we discuss in Section~\ref{sec:coop-discussion}.

\subsubsection{Case presentation}\label{sec:coop-cases-pres}
We present the cases below (see summary information in Table~\ref{tab:project-summary}).

\textit{TensorFlow by Google:} TensorFlow is an open source deep learning framework that is widely used in academia and industry for creating and training machine learning models \cite{google_TensorFlow_2023}. It was started by Google Brain in 2011 to facilitate the use of neural networks in Google research and products \cite{abadi_tensorflow_2016}. TensorFlow was publicly released in 2015, and TensorFlow 2 was released in 2019.  After its initial release, Jeff Dean from Google stated, ``We’re hoping that the community adopts this as a good way of expressing machine learning algorithms of lots of different types and contributes to building and improving [TensorFlow] in lots of different and interesting ways'' \cite{metz_google_2015}. Other reported incentives include increasing adoption, benefiting from crowdsourced innovation, and recruitment \cite{metz_google_2015-1}. 

\textit{PyTorch by Meta:} PyTorch is an open source deep learning framework widely used in academia and industry for training neural networks \cite{PyTorch_PyTorch_2023}. It was released in 2016 and maintained by Facebook AI Research at Meta until its donation to the LF's PyTorch Foundation in September 2022~\cite{zemlin_welcoming_2022}. Mark Zuckerberg, Meta's CEO, has spoken publicly about the benefits that Meta has derived from the popularity of PyTorch in AI R\&D, in particular the crowdsourced innovations that developers in the wider PyTorch ecosystem have contributed back to PyTorch, resulting in improvements \cite{south_park_commons_mark_2024}.

\textit{Transformers by HF:} Transformers provides APIs and tools to download, fine-tune, and share pre-trained machine learning models hosted on the HF Hub~\cite{huggingface_transformers_2023}. Transformers is integrated with TensorFlow and PyTorch, but operates at a higher level of the AI stack. It is developed by HF, a start-up, whose mission is to democratise AI by providing accessible tools and resources for researchers and developers. While HF initially developed a chatbot app, it is now better known for the HF Hub, which hosts a fast-growing number of pre-trained models and datasets, and its OSS libraries (e.g., Transformers and Diffusers), which allow researchers and developers to download, modify, and share models and datasets hosted on the HF Hub. In light of the emerging popularity of its tools and the HF Hub, the start-up has raised hundreds of millions of US dollars in investment~\cite{dillet_hugging_2022}. 

\subsection{Software repository mining}\label{subsec:data-dm}
\subsubsection{Data collection}
We mined data from each repository's commit logs on GitHub in order to analyse code authorship patterns and distributions. Specifically, we obtained historical commit data from each repository via the GitHub REST API, spanning from the date of the first commit in each repository until 12 September 2022, which we set as the data collection cut-off date to count PyTorch as a company-hosted project. Each commit dataset includes per commit: \texttt{sha},  \texttt{date},  \texttt{name},  \texttt{email address},  \texttt{modified source files}, and \texttt{lines of code (LOC)} added,  \texttt{LOC deleted}, and  \texttt{LOC changed (net)}. We acknowledge that while this data collection cut-off date predates major developments in the AI industry, it is defensible given that our study focuses on testing and extending theory on open source co-opetition in company-hosted OSS projects with insights from the AI industry, rather than focusing on trends in the AI industry per se. We discuss this temporal limitation in in Section~\ref{sec:coop-threatstovalidity}.

\newpage
\vspace*{\fill}
\begin{table}[ht]
\centering
\small
\caption{Summary information about TensorFlow, PyTorch, and Transformers}
\begin{tabular}{p{3.5cm}>{\centering\arraybackslash}p{3.1cm}>{\centering\arraybackslash}p{3.1cm}>{\centering\arraybackslash}p{3.1cm}}
\toprule
\textbf{} & \textbf{TensorFlow} & \textbf{PyTorch} & \textbf{Transformers} \\ \midrule
\textbf{Project Ownership} & ~ & ~ & ~ \\
Released by & Google & Meta & HF \\
Repository owner & Google & Meta & HF \\
GitHub organisation & TensorFlow & PyTorch & HF \\ \midrule
\textbf{Company Information} & ~ & ~ & ~ \\
Company size & Large & Large & Small \\
Size (employees) & \textasciitilde190,000 \cite{statista_alphabet_2022} & \textasciitilde86,000 \cite{statista_meta_2022} & \textasciitilde120 \cite{perez_hugging_2022} \\
Market valuation (USD) & \textasciitilde1.4T \cite{companies_market_cap_alphabet_2023} & \textasciitilde450B \cite{companies_market_cap_meta_2023} & \textasciitilde2B \cite{dillet_hugging_2022} \\
OSS criticality for business & N & N & Y \\ \midrule
\textbf{Project Information} & ~ & ~ & ~ \\
Initial release & November 2015 & September 2016 & November 2018 \\
OSS license & Apache 2.0 & BSD 3-Clause & Apache 2.0 \\
\# contributors & 3,197 & 2,430 & 1,392 \\
\# GitHub stars & 168,000 & 58,600 & 70,000 \\
\# commits & 135,051 & 51,538 & 10,609 \\
\# forks & 87,200 & 16,300 & 16,100 \\ \midrule
\textbf{Resourcing \newline by Company} & ~ & ~ & ~ \\
Employs maintainers & Y & Y & Y \\
Organises events & Y & Y & Y \\
Produces courses/tutorials & Y & Y & Y \\ \midrule
\textbf{Legal Control \newline by Company} & ~ & ~ & ~ \\
Enforces CLA & Y & Y & N \\
Owns trademark & Y & Y & Y \\ \midrule
\textbf{Contribution Policy \newline by Company} & ~ & ~ & ~ \\
Sets contribution policy & Y & Y & Y \\
Resolves CoC violations & Y & Y & Y \\
Encourages docs fixes & Y & Y & Y \\
Encourages issue reports & Y & Y & Y \\
Recommends first issues & Y & Y & Y \\
Encourages PRs & Y & Y & Y \\
Encourages PR reviews & Y & Y & NaN \\ \midrule
\textbf{Company Branding} & ~ & ~ & ~ \\
On documentation & N & N & Y \\
On project website & N & N & Y \\
\bottomrule
\end{tabular}
\label{tab:project-summary}
\\
{\small\textit{N.B.: Project information as of 12 September 2022.}}
\end{table}
\vspace*{\fill}
\newpage

\subsubsection{Username merging}
We merged multiple identities for unique developers, which is a common problem in software engineering research that arises when developers use multiple accounts on GitHub or due to how Git records their local credentials \cite{bird_mining_2006,goeminne_comparison_2013,kouters_whos_2012,robles_developer_2005}. Following prior work \cite{zhu_empirical_2019}, we built two bipartite networks that respectively mapped each username to all previously used corresponding email addresses and each email address to all previously used corresponding usernames. Then, we merged identities based on the linked username-email address pairs and email address-username pairs. For analytical purposes, we created a unique user ID for each developer identity. This resulted in 3,434$\rightarrow$3,058, 3,964$\rightarrow$3,564, and 1,479$\rightarrow$1,392 contributors in the PyTorch, TensorFlow, and Transformers datasets respectively. Three reviewers cross-validated the accuracy of this approach, identifying 6, 3, and 2 errors in respective datasets.

\subsubsection{Bots removal}
Following prior work~\cite{zhang_how_2020,robles_developer_2005, lin_developer_2017}, we removed commits by bots from the datasets. Specifically, we dropped 919 (1.78\%), 38,503 (28.20\%), and 23 (0.22\%) commits from the PyTorch, TensorFlow, and Transformers datasets respectively.

\subsubsection{Affiliation identification}
Following prior work \cite{zhang_how_2020,zhang_corporate_2022}, we applied a semi-automated approach to identify the affiliations of contributors at the time of each commit. We mined affiliations from the email addresses associated with each commit, which is considered the most accurate source of affiliation data \cite{nguyen-duc_software_2019,mehra_firms_2011}. The affiliations of commits with consumer email addresses, identified using a publicly available list \cite{ihmpavel_free-email-domains-listsrcconstantsts_2022,valiev_ecosystem-level_2018}, or no email addresses, were left blank. This identified the affiliations of 37.0\%, 48.9\%, and 9.2\% of contributors in the PyTorch, TensorFlow, and Transformers datasets respectively. We discuss the divergence in affiliation identification per project in Section~\ref{sec:coop-threatstovalidity}. 

For contributors with missing affiliations, we mined affiliations from users' GitHub profiles. To address data quality concerns (e.g., self-reported affiliations are not time-sensitive to activity), three authors cross-validated affiliation data for contributors with 5 or more commits, and manually labelled missing values through Internet searches. We filtered out contributors who had submitted less than 5 commits to limit the analysis to contributors who met a minimum activity threshold and to reduce data labelling burden. Contributors who did not work for a company were recorded as ``volunteers'' and unidentifiable affiliations were recorded as ``unknown.'' When contributors used both company and private email addresses, we linked all commits to their company affiliation. 

Three authors reviewed 100 randomly sampled commits from each project to estimate agreement in the manual labelling. We found 5 inconsistencies, indicating 98.3\% agreement. We could not cross-validate the entire dataset  due to resource constraints, as finding the affiliation(s) of one contributor took up to 10 minutes. We further evaluated the accuracy of the automated approach against the manually validated ground truth, finding the automated approach had labelled 83.9\%, 92.0\%, and 39.5\% of commits correctly for PyTorch, TensorFlow, and Transformers respectively.

\subsubsection{Descriptive analysis}
We report the provenance of commits per affiliation type in Figure~\ref{fig:coop-commit-provenance}; the dominance of host companies in Table~\ref{tab:coop-project-dominance}; and the relative contributions of the top ten companies per project in Tables~\ref{tab:coop-pytorch-top-companies}-\ref{tab:coop-transformers-top-companies}.

\subsection{Social network analysis}\label{subsec:data-sna}
 
\subsubsection{Operationalising collaboration}
We employed SNA to investigate open source co-opetition, drawing on prior prior that used SNA to study collaboration among individual developers  \cite{crowston_social_2005,bird_mining_2006,madey_open_2002,singh_small-world_2010,tan_social_2007} and companies \cite{linaker_how_2016,linaker_method_2020,snarby_collaboration_2013,teixeira_cooperation_2016,zhang_how_2020}. While some prior work has operationalise collaboration as discussions in issue trackers \cite{nguyen-duc_software_2019, teixeira_cooperation_2016, linaker_how_2016}, we operationalised collaboration as commits made by a pair of contributors to the same file during a release cycle. We did this for two reasons. First, commits represent an accurate, timestamped audit trail of code authorship \cite{orucevic-alagic_network_2014, zhang_how_2020}; and second, examining interactions on code authorship per release enables longitudinal analysis of collaborations \cite{basole_visualization_2009,teixeira_cooperation_2016,linaker_how_2016}. The second step concerned the choice of the unit of analysis. Since this study is concerned with co-opetition between companies, we chose the company as the unit of analysis, following prior work \cite{nguyen_duc_coopetition_2017,teixeira_cooperation_2016,zhang_how_2020}. However, we acknowledge that the aggregation to the company level loses crucial information about the activity of individual developers \cite{dahlander_man_2006,zhou_developer_2010}.

\subsubsection{Network construction}
We recorded directed edges between pairs of contributors, who had contributed to the same file(s) during a release cycle, with edges weights corresponding to LOC changed in said file(s) by the respective contributor. For example, if contributor A modified 5 LOC in file F and contributor B modified 6 LOC in file F during the same release, the edge weights for A->B and B->A would be 5 and 6 respectively. 
The directed networks can therefore be formally represented as \(G = (C, A_c, E, W_{ij})\), where \(C\) is the set of developers, \(E\) is the set of edges, \(A_c\) is the set of node attributes, and \(W_{ij}\) is the edge weight. We aggregated the release networks into annual snapshots to enable comparative analysis \cite{long_social_2007}. Next, we assigned unique user IDs to nodes and excluded bots to limit the network to human-to-human collaboration. Then, we constructed company networks by merging developer nodes with the same affiliation, combining their edges, and summing their edge weights. We filtered out nodes with ``volunteer'' or ``unknown'' affiliations.

\subsubsection{Network analysis}
We performed the network analysis in three steps. First, we measured three kinds of network centrality to understand different aspects about companies' roles in the collaboration networks (see Tables~\ref{tab:coop-pytorch-top-companies}-\ref{tab:coop-transformers-top-companies}). Specifically, out-degree indicates a company's breadth of collaborations \cite{networkx_out_degree_centrality_2023}, PageRank suggests its global importance \cite{networkx_pagerank_2023}, and betweenness centrality reflects its brokerage role \cite{networkx_betweenness_2023}. Second, we visualised annual network snapshots to observe changes in the collaboration relationships between companies \cite{teixeira_collaboration_2014} and the role of individual companies \cite{linaker_method_2020}. For readability, we filtered the networks to the 20 nodes with the highest degree centrality (Figure~\ref{fig:coop-network-vis}). Third, to account for network size effects in the former steps, we analysed three size-independent metrics of the complete networks over time (see Tables~\ref{tab:network-metrics-pytorch}-\ref{tab:network-metrics-transformers}). Specifically, degree centralisation measures how much the network structure is organised around focal nodes \cite{freeman_centrality_1978}; degree skew indicates the asymmetry of the degree distribution, helping to identify the presence of hubs \cite{barabasi_emergence_1999}; and the clustering coefficient quantifies the tendency of nodes to cluster together, helping to identify the presence of tightly-knit communities of collaborating companies \cite{watts_collective_1998}.

\subsection{Semi-structured interviews}\label{sec:coop-data-interviews}

\subsubsection{Interviewee sampling}
We recruited 10 company-affiliated contributors for interviews (see Table~\ref{tab:coop-respondent-list}). Our sampling approach involved sending interview invitations to a subset of company-affiliated developers, who were not affiliated with the respective host company \cite{nguyen-duc_software_2019}. In exchange for their time, we offered to donate 15 USD to a project of their choice. In total, we sent 350 emails to 150 TensorFlow contributors, 150 PyTorch contributors, and 50 Transformers contributors. We sent fewer emails to Transformers contributors due to less company-affiliated contributors in this project. We received 13 responses (3.71\%) and 10 acceptances (2.86\%).

\subsubsection{Semi-structured interviews}
We conducted 10 digital, semi-structured interviews, which lasted between 30 and 60 minutes. The semi-structured interviews followed an interview guide with five topics: their personal and employer's incentives; their individual and employer's contribution strategies, if any; their experience of collaborating with developers employed by the host company and/or other companies; a discussion of the quantitative findings; and their views on the unique aspects of open source co-opetition in company-hosted OSS projects. During the interviews, we showed the network visualisations to the respondents in order to elicit responses about the evolving relationships between companies in the three projects \cite{molina_giving_2014,hogan_visualizing_2007,tubaro_visual_2016}. Specifically, we asked respondents to identify collaborations between companies that they were aware of, to explain their understanding of the nature and incentives for these collaborations, and to comment on changes in collaborations between companies as shown in the network visualisations (see Figure~\ref{fig:coop-network-vis}).

\begin{table}[t]
	\centering
	\footnotesize
	\caption{List of respondents and their affiliations}
	\begin{tabular}{p{0.5cm}p{3cm}p{3cm}p{5cm}}
	\toprule
    	\textbf{ID} & \textbf{Sector} & \textbf{Company Size} & \textbf{Project(s)} \\ \midrule
    	A & Software consultancy & Small & PyTorch \\
    	B & Transportation, IT & Large, small & PyTorch, TensorFlow, Transformers \\
    	C & IT & Large & TensorFlow \\
    	D & IT & Medium & PyTorch \\
    	E & Transportation, IT & Large & TensorFlow \\
    	F & Software consultancy, IT  & Small, large & PyTorch \\
    	G & E-commerce, IT & Small, small & PyTorch, Transformers \\
    	H &  Software consultancy & Small & PyTorch \\
    	I & IT & Medium & TensorFlow \\
    	J &  Software consultancy & Small, medium & PyTorch \\ \bottomrule
	\end{tabular}
	\label{tab:coop-respondent-list}
\end{table}

\subsubsection{Thematic analysis}
We analysed the interview data following a systematic six-step procedure for thematic analysis; that is, the identification, analysis, and reporting of themes in qualitative data \cite{braun_using_2006}. We adopted an integrated approach to code and identify themes in the interview data, combining deductive and inductive methods \cite{cruzes_recommended_2011}. In particular, we used key findings from prior work as initial categories for the deductive coding, whilst inductively coding the interview data following grounded theory approaches to capture novel themes \cite{charmaz_constructing_2006}. This combination enabled us to both test prior theory and uncover new findings. The first author performed the initial coding until reaching saturation \cite{charmaz_constructing_2006}. A second author validated the codes to enhance the reliability of the analysis, and subsequently the codes were merged into themes \cite{lincoln_naturalistic_1985}. Finally, we member-checked themes with respondents to ensure accuracy and practical relevance \cite{lincoln_naturalistic_1985}. When we quote respondents in Section~\ref{sec:coop-results}, we identify them with their ID from Table~\ref{tab:coop-respondent-list} and mention their project(s) in abbreviated form in brackets (PT for PyTorch, TF for TensorFlow, and TR for Transformers). 

\section{Results} \label{sec:coop-results}

\subsection{RQ1: What, if any, are typical patterns of open source co-opetition in company-hosted OSS projects?}

\vspace{1em}

\begin{tcolorbox}[colback=yellow!20, colframe=yellow!80, title=\textbf{\textcolor{black}{Key findings}}]
The three projects reveal similar patterns of code authorship between host and external companies, yet distinct structures of collaboration. In each project, the host and external companies account for $\sim$80\% and 20\% of commits respectively. PyTorch and TensorFlow have decentralised network structures with lower degree centralisation, lower degree skew, and higher clustering coefficients, indicating strong inter-connections between companies. By contrast, Transformers has a hub-and-spoke network structure with higher degree centralisation and lower clustering, underlining HF's broker role between external companies.
\end{tcolorbox}

\vspace{1em}

\subsubsection{Distribution of code authorship by host and external companies}
\begin{figure}[ht]
  \centering
  \includegraphics[width=\linewidth]{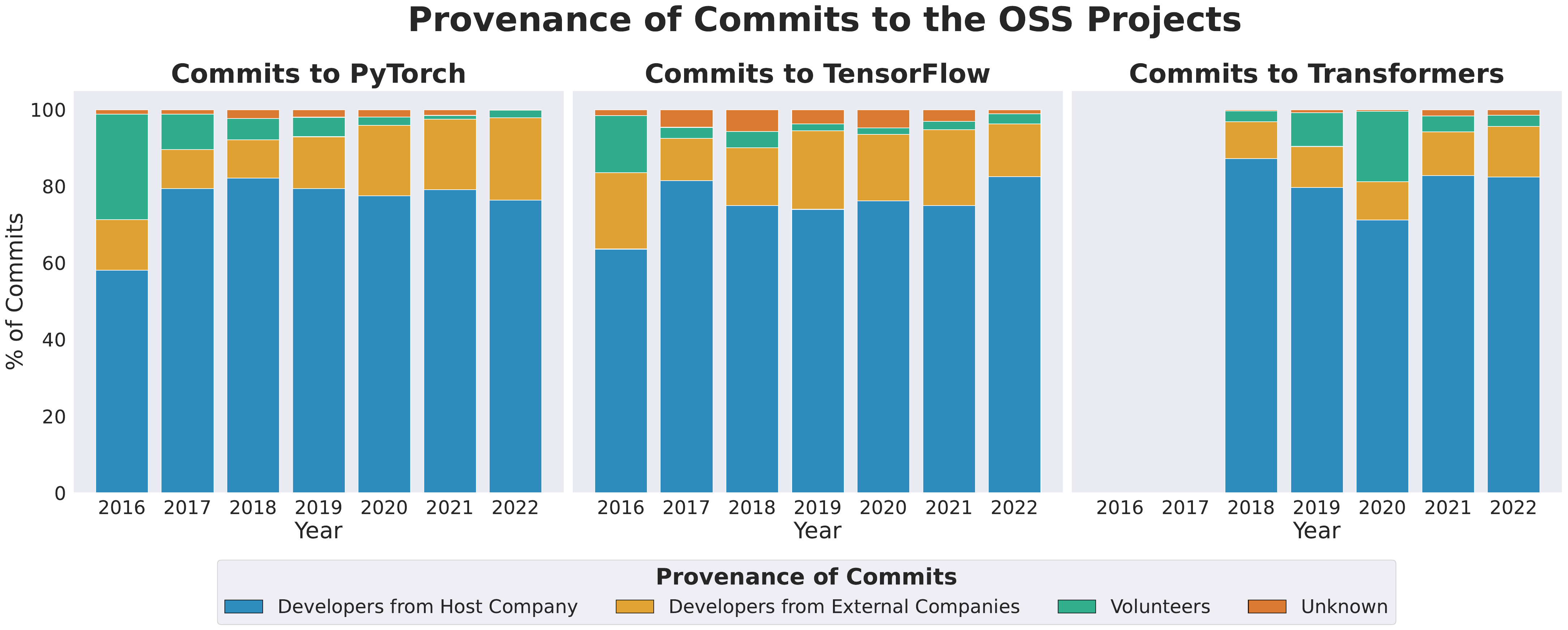}
  \caption{Provenance of commits to OSS projects per year}
  \label{fig:coop-commit-provenance}
\end{figure}

Host companies are dominant in their respective projects by several metrics (see Table~\ref{tab:coop-project-dominance}). Meta, Google, and HF employ 61.25\%, 47.61\%, and 32.18\% of contributors to their respective projects. These percentages increase in the maximal \textit{k}-cores of the annual network snapshots, indicating the host companies' control over core development. For example, in the 2022 network snapshots, 31\%, 50\%, and 9\% of contributors to PyTorch, TensorFlow, and Transformers respectively were affiliated to the host company, rising to 42\%, 68\%, and 38\% in the network cores. Host company employees account for approximately 80\% of annual commits, while external companies contribute 10-20\% of annual commits (Figure~\ref{fig:coop-commit-provenance}). The Pareto principle is evident in each project, with less than 20\% of contributors responsible for more than 80\% of commits. Transformers has the most imbalanced authorship, with 7.54\% of contributors making 80\% of commits, and has a low bus factor due to most commit activity coming from a few highly active contributors.

\vspace{1em}

\subsubsection{Collaboration Networks: Centralised vs Decentralised Collaboration}

We observe distinct patterns of collaboration across the three projects, with PyTorch and TensorFlow exhibiting decentralised structures despite dominant code authorship by the host companies, while Transformers shows a hub-and-spoke structure. In PyTorch, while Meta contributes the majority of commits (84\%) and lines of code (84\%), many commits are from Nvidia, Intel, AMD, and Google (see Table~\ref{tab:coop-pytorch-top-companies}). Similarly, in TensorFlow, Google dominates in commits (85\%) but contributes a smaller share of lines of code (34\%), with significant contributions from Nvidia, Intel, and IBM, among others (see Table~\ref{tab:coop-tensorflow-top-companies}). Despite this concentration of authorship, external companies have high out-degree centrality values in both projects, indicating active collaboration on project files among these companies. However, Transformers presents a contrasting picture. HF not only dominates in code authorship (91\% of commits, 94\% of lines of code) but also in network centrality measures (see Table~\ref{tab:coop-transformers-top-companies}). External companies have low out-degree and PageRank centrality, indicating limited breadth of collaborations on project files and their global network importance, while HF has high betweenness centrality, indicating its pivotal role as a broker between external companies.

\begin{figure}[htbp]
  \centering  

	\begin{subfigure}{\linewidth}
	\centering
	\includegraphics[width=\linewidth, height=6cm]{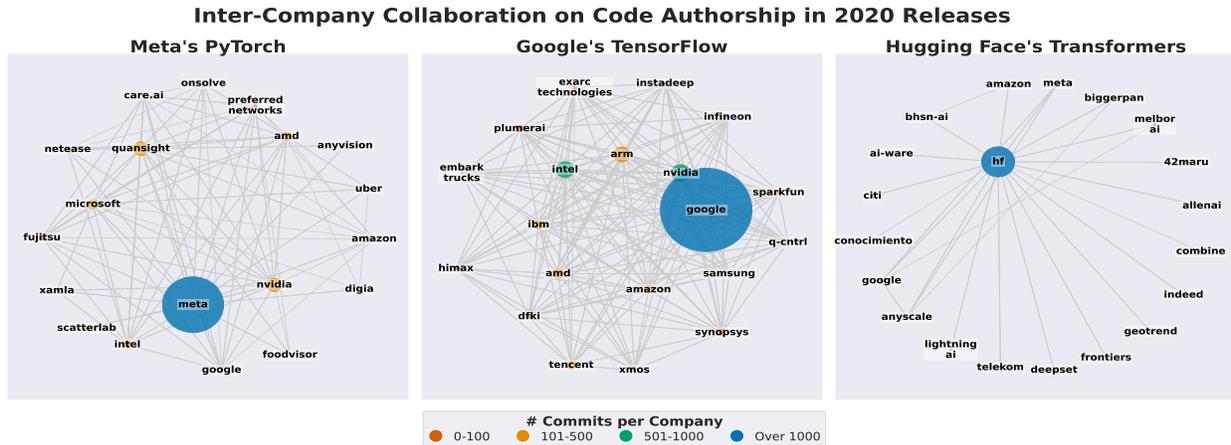}
	\caption{Inter-Company Collaboration on Code Authorship in Project Files in 2020 Releases}
	\vspace{10pt}
  \end{subfigure}

  \vspace{1em}
 
  \begin{subfigure}{\linewidth}
	\centering
	\includegraphics[width=\linewidth, height=6cm]{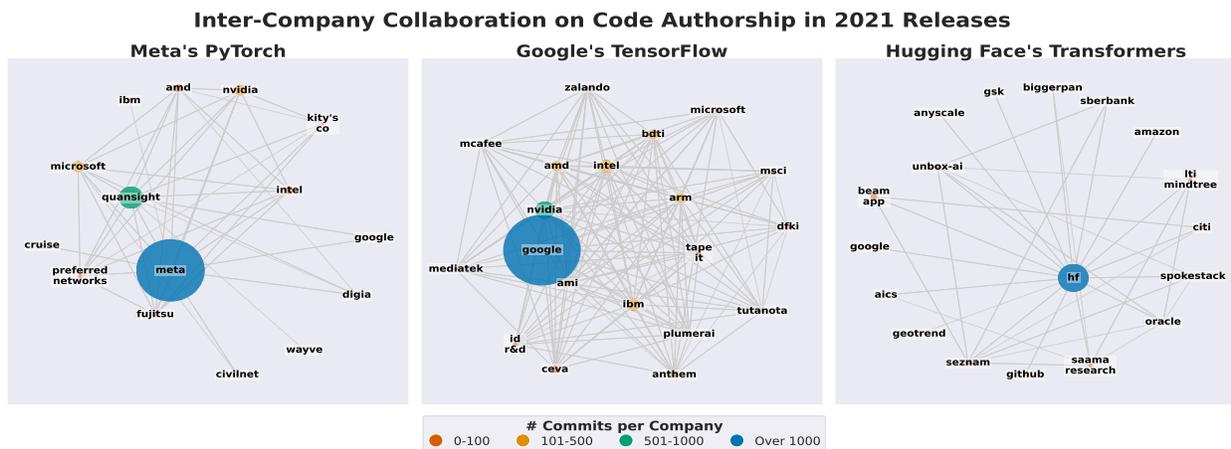}
	\caption{Inter-Company Collaboration on Code Authorship in Project Files in 2021 Releases}
	\vspace{10pt}
  \end{subfigure}

    \vspace{1em}
   
  \begin{subfigure}{\linewidth}
	\centering
	\includegraphics[width=\linewidth, height=6cm]{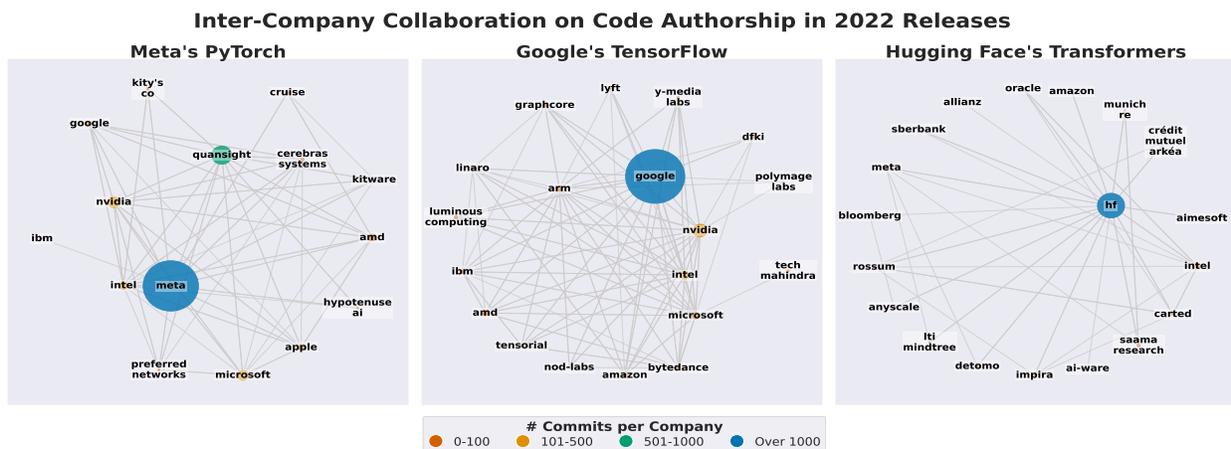}
	\caption{Inter-Company Collaboration on Code Authorship in Project Files in 2022 Releases}
	\vspace{10pt}
  \end{subfigure}

  \caption{Open Source Co-opetition Networks in the PyTorch, TensorFlow, and Transformers Projects}
  \label{fig:coop-network-vis}
\end{figure}

The network visualisations underscore these structural differences (see Figure~\ref{fig:coop-network-vis}). PyTorch and TensorFlow display decentralised collaboration with dense collaboration between various companies, while Transformers exhibits a hub-and-spoke structure, with HF playing a broker role between peripheral companies. These differences persist despite variations in network size across projects and over time (see Tables~\ref{tab:network-metrics-pytorch}-\ref{tab:network-metrics-transformers}), corroborating the observed differences in Figure~\ref{fig:coop-network-vis}. For example, the Transformers networks exhibit higher degree centralisation and degree distribution skew compared to PyTorch and TensorFlow, indicating a collaboration structure organised around a focal company (i.e., HF). By contrast, the PyTorch and TensorFlow networks have higher clustering coefficients, suggesting the presence of more decentralised, interconnected communities of companies that collaborate on code authorship in these projects.

\subsection{RQ2: What types of collaborative relationships do companies pursue in company-hosted OSS projects, and why?}

\begin{tcolorbox}[colback=yellow!20, colframe=yellow!80, title=\textbf{\textcolor{black}{Key findings}}]
Host and external companies collaborate in, at least, three types of relationships in company-hosted OSS projects: strategic, contractual, and non-strategic collaborations. Strategic collaborations are primarily dyadic relationships between the host company and an external company, often characterised by private collaborations and driven by competitive incentives. Contractual collaborations involve the outsourcing of development to third-party companies. Non-strategic collaborations encompass a range of contribution types and motivations, such as hobbyism, bug-fixing, code adoption, and corporate OSS initiatives, which blur the line between voluntary and work-based contributions by company-affiliated contributors.
\end{tcolorbox}

\subsubsection{Strategic collaborations}\label{sec:coop-results-strategic-collabs}
The first type of inter-company collaboration that we observe concerns strategic collaborations between the host company and an external company. In each of the three projects, the host company has engaged in strategic collaborations with external companies, such as AI accelerator manufacturers, cloud service providers, and AI model producers, where business objectives and competitive dynamics have been at play. Contrary to prior work, we find that business strategy plays an important role at the developer level in such scenarios. For example, competing AI accelerator manufacturers and cloud service providers pursue strategic collaborations with Meta and Google to ensure compatibility between new releases of the frameworks and their respective hardware or cloud services. Respondent F (PT) explained that their company had clear goals on what they wanted to achieve in PyTorch, and their team's contributions were ``narrowed down'' accordingly. In regular closed-door meetings with the PyTorch maintainers, which were held privately for select company-affiliated developers to protect proprietary information, they would exchange information about their upcoming releases and contribution priorities. Since it was also in the interest of Meta to have PyTorch run efficiently on their AI accelerators, they contended that this relationship could be characterised as a strategic interdependence.

Competitive logics between market rivals are key to these strategic collaborations. Respondent B (PT, TF, TR) explained that it is a strategic priority for the AI accelerator manufacturers to minimise the risk of being undercut by a market rival if these popular deep learning frameworks do not run efficiently on their hardware. Respondent F (PT) agreed, explaining, ``When you’re contributing to PyTorch on behalf of the company, you have to think about how it makes profit for the company. So, our very first goal was to make sure that PyTorch works well on our GPUs.'' Similarly, respondents discussed the self-interest of Google to collaborate with AI accelerator manufacturers, despite also being in the business of selling AI accelerators, specifically Tensor Processing Units, to ensure optimal performance of TensorFlow on other hardware offerings and vice versa, especially ``Nvidia’s GPUs which are the market leader'', in order to avoid being undercut by PyTorch. Commenting on Figure~\ref{fig:coop-network-vis}, Respondent D (PT) asserted that these strategic collaborations are ultimately about downstream revenue generation: ``AWS have an incentive because everyone wants to buy their servers to train their models and people are buying Nvidia's cards to train their neural networks. So, their incentive is to indirectly boost their sales by helping make PyTorch run really well.''  Contrary to prior work \cite{nguyen-duc_software_2019}, business strategy evidently influences the activity of developers who contribute to projects on behalf of their employers.

While Figure~\ref{fig:coop-network-vis} does not visualise comparable evidence of open source co-opetition in Transformers due to the data collection cut-off in 2022, Respondent G (PT, TR) explained that market rivals in the cloud compute market like Microsoft Azure, AWS, and Google Cloud had since become active contributors to the project, as maintainers of integrations with their respective cloud offerings. Furthermore, they explained that HF engages in strategic, closed-door collaborations with AI companies like Meta and Stability AI to offer day-zero integrations for their open model releases. They explained, ``That’s something we need to be active upon, we can’t really delegate that to the community to be able to, you know, quickly sprint on it and quickly deliver because there’s a lot of advantages of being the first movers here.'' The existence of these private collaborations creates a layer of strategic interaction that exists alongside, but separate from, the broader community. As a result, it can be difficult for developers that are not affiliated with one of these companies to participate in such strategic collaborations. For example, Respondent F (PT) explained they had found it difficult to contribute to issues or PRs related to request for comment (RFC) documents submitted by AI accelerator manufacturers since they referred to undisclosed proprietary information. They commented, ``There are a lot of things that you will just not know as an outsider.'' However, Respondent C (TF) explained that due to the interdependence between the deep learning frameworks and AI accelerators, third-party companies do collaborate with the corresponding companies in some way, albeit without access to their private communications or meetings, such as by reporting issues and submitting PRs on hardware-related features.

\subsubsection{Contractual collaborations}
The second type of inter-company collaboration that we observe concerns contractual collaborations between the host company and an external company, to which the host company outsources development and maintenance tasks. In PyTorch, Meta outsources development to QuanSight, a consultancy that focuses on the Python data science stack. After Meta, QuanSight has made the most contributions to PyTorch (see Table~\ref{tab:coop-pytorch-top-companies}). Respondent A (PT) explained that their team contributes to a range of core modules in PyTorch: ``We’re not helping them build a particular product, we’re just helping them build PyTorch as a project.'' Due to the contractual nature of the collaboration between Meta and QuanSight, QuanSight developers support Meta developers with the realisation of their strategic goals in PyTorch, ``which are essentially for PyTorch to be cutting edge and performing well across the board.'' Respondent H (PT) speculated that Meta had contracted this work out to leverage their OSS know-how and to reduce costs since outsourcing was cheaper than paying the salaries of Meta engineers.

While the contractual nature of the relationship between Meta and QuanSight involves closed-door communication and collaboration, Respondent A (PT) commented that Meta grants QuanSight developers the freedom to contribute as if they were organic contributors. Since their mandate is not focused on specific features, they collaborate with all kinds of contributors, including volunteers and company-affiliated contributors. They explained that QuanSight contributors mostly work independently: ``Maybe once or twice a week, they’ll check in and report up to the project managers what they’ve been doing, but essentially it’s up to them to take the initiative to find higher priority issues and just make sure that they get resolved.'' They added that they also deploy small teams of developers to work with a lead engineer from Meta, who gives them direction. Given how closely QuanSight and Meta employees collaborate in PyTorch, they suggested that it can be hard to distinguish between Meta and QuanSight employees. ``You know, if you were to sit over their shoulder and watch them day to day without any idea of what company they work for, you would probably just assume that they were all one big team.'' This contractual collaboration exemplifies a hybrid model, where contracted developers are granted autonomy while still needing to align with the host company's goals, thus blurring the lines between internal and external contributors.

\subsubsection{Non-strategic collaborations}
The third type of inter-company collaboration that we observe concerns non-strategic collaborations between the host company and developers from external companies, who have a number of incentives for contributing to company-hosted OSS projects, including their personal interest, bug-fixing, and OSS contribution initiatives at work. For example, Respondent F (PT) explained they got involved in PyTorch following a summer internship at an academic lab, which sparked their interest in deep learning and led them to ``dive deeper into the PyTorch code base''. 
Meanwhile, four respondents explained that they had contributed to projects to fix bugs they encountered at work. However, Respondent B (PT, TF, TR) explained that while they made fixes to TensorFlow for work purposes, they underlined that there was no direction from their employer. They contrasted these contributions with strategic contributions their start-up now makes to PyTorch when they encounter issues that are specific to the company’s use cases and not ``in the top of priority list for the PyTorch maintainers.'' 

The lines are blurred about whether such contributions should be considered as voluntary or company-affiliated contributions. In fact, the distinction between voluntary and company-affiliated contributions was not a relevant consideration for several respondents, who had engaged in such non-strategic collaborations. For example, Respondent D (PT) explained that even though they use their company-affiliated GitHub account for all of their activity, they did not attribute this activity to their company because ``OSS is a hobby of mine.'' Their principal contribution to PyTorch was the improvement of its implementation of xdoctests, a Python package for executing tests in documentation strings that they maintain. Since then, they explained that PyTorch maintainers reach out from time to time to get their advice on related problems, which they help out which on a voluntary basis. They explained, ``If I’m not working on a project or something explicitly beneficial for my company, I don’t charge my time.'' Cases as such demonstrate the importance of hobbyism among company-affiliated developers who contribute to company-hosted OSS projects.

Corporate OSS contribution initiatives and personal code adoption goals were mentioned as additional non-strategic reasons for contributions to the company-hosted OSS projects. Respondent C (TF) explained that they began contributing to TensorFlow through an OSS contribution initiative by their employer, which ``started out as a side hobby'' but they got more involved when a maintainer invited them to collaborate on API migrations for tf.data. They used a restricted-access Google document with the list of APIs to migrate, but ``the conversation was mostly through PRs on GitHub.'' While they contributed during working hours, they received no direction from their employer. Meanwhile, Respondent B (PT, TF, TR) contributed a modified version of research code, initially developed for a scientific paper and published with a non-commercial use license, to the Transformers repository to increase its adoption and impact. These non-strategic collaborations highlight the complex interplay between personal interests, professional development, and corporate initiatives in driving contributions to company-hosted OSS projects, often transcending clear-cut categorisations of voluntary or company-affiliated work.

\subsection{RQ3: What similarities and differences characterise open source co-opetition in company-hosted OSS projects compared to foundation-hosted OSS projects?}

\begin{tcolorbox}[colback=yellow!20, colframe=yellow!80, title=\textbf{\textcolor{black}{Key findings}}]
There are both similarities and differences in open source co-opetition in company-hosted OSS projects compared to foundation-hosted OSS projects, as identified by prior work. On the one hand, prior findings about the role of gatekeepers, the competitive use of forks, and non-competitive attitudes among developers apply to a certain extent in the three projects. On the other hand, single-vendor governance introduces a power imbalance that affects open source co-opetition practices and possibilities in company-hosted OSS projects, from the risks of a host company's unilateral decision-making power (e.g., to change the license) to their community involvement strategy (e.g., from over-control to over-delegation).
\end{tcolorbox}

\subsubsection{The limited role of gatekeepers}
We observe that developers perform gatekeeper roles to coordinate contributions made on behalf of their company in some but not all cases in the three projects. For example, in strategic collaborations, lead developers or engineering managers coordinate closed-door meetings and information sharing. However, respondents contested the universality of gatekeepers. Respondent F (PT) asserted, ``There are always gatekeepers, right? The PyTorch team was just like any other team at [company], we always had a tech lead, we always had a manager, it’s not like we were operating independently. They smoothen the communication between the two companies.'' They explained that while engineering managers were key coordinators and communicators in their biweekly meetings with Meta, junior developers were given ample opportunity to communicate their views ``and they were heard''. The gatekeeper role was even less important in contractual and non-strategic collaborations. For example, Respondent C (TF) ``did not have a specific direction or requirements from [company]'' during their contributions to TensorFlow, while Respondent H (PT) enjoyed freedom to work on wide-ranging PyTorch issues due to QuanSight's contractual agreement with Meta. These findings complicate prior theory on the indispensable role of gatekeepers in facilitating open source co-opetition, providing critical nuance on the relevance of gatekeepers in different types of collaborations \cite{nguyen_duc_coopetition_2017,nguyen_duc_software_2019}.

\subsubsection{Attitudes towards ``competitors''}
We find that prior findings about non-competitive attitudes among company-affiliated developers apply to a certain extent in the three projects \cite{nguyen_duc_coopetition_2017,nguyen-duc_software_2019}. Respondents, who had participated in non-strategic and contractual collaborations, did not view contributors from other companies, including market rivals, as competitors. Respondent H (PT) commented that the affiliation of contributors was not important to them, citing their philosophy that ``everyone grows during collaboration, but not everyone grows during competition.'' Respondent C (TF) commented that their manager was not concerned that they were contributing to a Google project: ``It was the other way around. They were appreciative that I was contributing to an open source project, which we used extensively.'' Furthermore, Respondent A (PT) argued that since PyTorch is a core library, they and their team do not have to navigate the competitive dynamics that a product team might face if they were to collaborate with a product team from another company. Several respondents suggested that competitive attitudes are the strongest when several market rivals are engaged in strategic collaborations with the host company at the same time. For example, Respondent F (PT) explained, ``There was always competition between AMD and Nvidia… if AMD had a bug because of which our task was slowing down, we would pitch to Meta that whatever AMD was doing, they [should] only touch their codebase and [not] affect our tasks and slow us down.'' These examples illustrate the presence of competitive attitudes among developers, thus contesting prior theory \cite{nguyen_duc_coopetition_2017,nguyen-duc_software_2019}.

\subsubsection{Different approaches to community involvement}
The respective host companies have taken different approaches to community involvement in their projects, which has shaped collaboration practices and possibilities. Numerous respondents remarked that Meta and HF take community-driven approaches. For example, Respondent B (PT, TF, TR) commended the ``back and forth'' discussions with PyTorch maintainers, and Respondent G (PT, TR) highlighted the open RFC process in Transformers where ``anyone from the community is encouraged to take part in the discussions.'' 
Respondents contrasted the friendliness of PyTorch and Transformers maintainers with their transactional interactions with TensorFlow maintainers. Respondent D (PT) commented, ``If I didn’t know that PyTorch was from Facebook, I don’t think I ever would have figured that out, whereas with Google it’s kind of obvious.'' However, community-oriented approaches can also backfire for host companies, when they fail to distribute power and decision-making in their projects. For example, Respondent B (PT, TF, TR) shared that they thought HF were ``not really espoused to open source philosophically, [rather] they’re espoused to it as a means to an end'' and they were ``reaping all the benefits'' from the community. They explained that when the start-up changed the license of its text-generation-inference repository without community deliberation, their company decided to stop contributing to HF's OSS projects to avoid such risks in the future. 

By contrast, respondents explained that Google takes a top-down approach to TensorFlow, with more obstacles for community involvement. For example, Respondent C (TF) explained, ``[Google] controls the level of granularity that other people should have access to. If there is something specific to Google, then they don’t release it.'' They also mentioned that Google develops TensorFlow privately through an internal RFC process, which is out-of-bounds to the broader contributor community. Similarly, Respondent E (TF) stated, ``I didn’t know what was on Google’s mind about what was most important and I certainly didn’t have a community feeling.'' Moreover, Respondent C (TF) explained that during their time collaborating with a TensorFlow maintainer on tf.data, they had limited awareness of the project’s overall roadmap: ``For tf.data, I understood what we were doing and why these migrations were necessary, but not the roadmap for TensorFlow as a whole. That is too complicated and not visible to outsiders.'' This lack of transparency, in turn, creates challenges for external contributors, who are blind to long-term roadmap priorities and may feel subordinate in the project's social hierarchy.

\subsubsection{Risk of license change}
Respondents underlined that a host company's ability to unilaterally change the license of its project without community deliberation is a key differentiating factor for open source co-opetition in company-hosted OSS projects. Respondent G (PT, TR) explained that one must be prepared for the event that the host company will change the license, while Respondent F (PT) commented that projects hosted by start-ups generally have higher risks than those hosted by industry giants because start-ups are under greater pressure to generate revenue from their OSS projects. Respondent B (PT, TF, TR) explained that HF had changed the license of its text-generation-inference project from Apache v.2 to a restrictive license, which was problematic for their commercial use of the project. He explained that in response a handful of companies, including his start-up, forked the project. While the fork allowed external companies to continue their collaboration on the project, they explained that ``this would never have happened at a foundation project''. Similarly, Respondent H (PT) remarked, ``the distribution of power is one thing that very much supports open source projects in general. However, if one company is in control, even if it’s managed nicely, you’re going to drive a lot of companies away just as a matter of principle or policy.'' Thus, the uneven playing field between host and external companies, including the risk of a sudden license change without deliberation with the wider contributor community, is a unique aspect of company-hosted OSS projects that influences open source co-opetition practices in such contexts.

\section{Discussion} \label{sec:coop-discussion}
In this section, we discuss the implications of our findings for research on open source co-opetition (see Section~\ref{sec:coop-impl-research}), as well as the threats to the validity of our findings (see Section~\ref{sec:coop-threatstovalidity}).

\subsection{Theoretical Contributions} \label{sec:coop-impl-research}

We set out to understand how companies collaborate on OSS development in the absence of vendor-neutral governance, specifically in OSS projects that are hosted and governed by one company. The findings from PyTorch, TensorFlow, and Transformers reveal collaboration strategies and challenges that both align with and diverge from prior theory on open source co-opetition.

\subsubsection{Host company dominance and divergent collaboration structures} 
The findings both confirm and contest findings from prior work. On the one hand, as expected, the host companies dominate code authorship in respective their OSS projects, contributing around 80\% of commits across the projects and over time, confirming prior work on commercial dominance \cite{zhou_inflow_2016, orucevic-alagic_network_2014}. On the other hand, our analysis of file-based collaboration reveals varying structures of inter-company collaboration in company-hosted OSS projects, from decentralised collaboration in PyTorch and TensorFlow to centralised collaboration in Transformers. The hub-and-spoke structure of collaboration observed in Transformers is particularly noteworthy, as HF publicly champions itself as ``the AI community building the future'' \cite{huggingface_hugging_2023}. However, we note that these variations simply reveal differences in how developers from various companies have interacted on project files through commits. These differences may reflect or be due to various project-specific factors, such as technical architecture, the host company’s contribution policy, or specific collaboration dynamics among companies. We encourage future research to investigate reasons for different inter-company collaboration structures to better understand this phenomenon.

\subsubsection{Advancing theory on open source co-opetition} 
Our findings extend prior theory on open source co-opetition by both testing prior theory and contributing novel findings from the context of company-hosted OSS projects. We find that prior theory regarding the key coordinating role of gatekeepers \cite{nguyen_duc_coopetition_2017,nguyen-duc_software_2019}, the strategic use of the fork as a competitive mechanism \cite{teixeira_collaboration_2014}, and non-competitive attitudes among individuals developers from rival companies \cite{nguyen_duc_coopetition_2017,nguyen-duc_software_2019} generalise to the three projects in our sample to a certain extent. However, our findings shed light on important nuances. First, we find that the coordinating role of the gatekeeper is not universally salient to open source co-opetition. While stakeholders like team managers play a coordinating role in strategic and contractual collaborations, such as by coordinating meetings between companies, respondents emphasised that they have enjoyed significant freedom when they have participated in such collaborations. Gatekeepers are even less relevant in non-strategic collaborations, where the lines are often blurred between voluntary and work-based contributions by company-affiliated developers. Moreover, contrary to prior findings \cite{nguyen-duc_software_2019}, we find that business strategy is indeed important at the developer level in some cases, in particular in strategic and contractual collaborations. What is more, while the respondents explained that in general they do not view contributors that are affiliated with market rivals as competitors \cite{nguyen-duc_software_2019}, competitive dynamics between rival companies, such as AI accelerator manufacturers, do influence interactions and priorities at the developer level. These findings contribute to a more nuanced understanding of open source co-opetition.

In addition, we contribute to open source co-opetition theory by characterising strategic, contractual, and non-strategic collaborations as three distinct collaborative relationship types between companies, which vary in terms of their collaboration practices and incentives. This typology builds upon the prior categorisation of intentional and passive collaborations between companies \cite{zhang_how_2020}, offering more granularity about the role of business strategy and collaboration practices in different types of collaborations between companies. In particular, our findings highlight that strategic collaborations, which often involve closed-door meetings and private communications between the host company and select external companies, create a private layer of interaction that exists alongside, but separate from, the broader community. This dual nature of collaboration---public and private--- presents unique challenges for outsiders. Developers who are not affiliated with collaborating companies find it difficult to contribute to certain issues or pull requests, particularly those related to proprietary information or undisclosed features. This dynamic highlights a potential tension between the open source ethos of transparency and the strategic needs of companies engaged in co-opetition. Future research should explore how this balance between openness and strategic privacy affects community dynamics, contributor motivation, and overall project sustainability. Furthermore, as the Meta-QuanSight relationship was the only observed contractual collaboration in our sample, further research is needed to validate the characteristics of contractual collaborations across a broader range of OSS projects, company profiles, and sectors. 

We also contribute novel insights into the unique aspects of open source co-opetition in the context of company-hosted OSS projects. In particular, single-vendor governance in company-hosted OSS projects creates a power imbalance between the host company and external contributors, leaving external contributors dependent and vulnerable to the decision-making and goodwill of the host company. Furthermore, we find that a company's approach to community involvement influences external companies' willingness to contribute, with both over-control and over-delegation potentially hindering collaboration. This extends prior work on the effects of commercial dominance \cite{zhou_inflow_2016,zhang_corporate_2022}, demonstrating that excessive task delegation, in addition to excessive dominance, can also deter contributors. While our findings advance understanding of open source co-opetition in company-hosted OSS projects, several questions remain. Since our findings are based on a limited sample of projects ($n=3$), further research is needed to validate these findings across a wider range of projects, companies, and sectors. Future research could also examine the impact of governance changes when projects transition to vendor-neutral foundations, providing timely insights into the relationship between project governance and collaboration dynamics between companies \cite{yue_igniting_2024,osborne_why_2024}.

\subsubsection{Open source co-opetition in the AI industry} 
The findings shed light on open source co-opetition dynamics specific to the AI industry, which manifest in, at least, three key areas of strategic collaboration. First, hardware-software optimization is a critical area of strategic collaboration, with Google and Meta engaging in strategic partnerships with AI accelerator manufacturers such as NVIDIA, Intel, and AMD. These collaborations aim to mutually optimise framework performance and hardware capabilities, even in cases where the companies are market competitors. Second, cloud integration is another area of strategic collaboration, with partnerships formed between framework maintainers and cloud service providers like AWS, Google Cloud, and Microsoft Azure. These collaborations seek to ensure seamless deployment and optimal performance of frameworks and AI models in cloud environments, reflecting the growing importance of cloud infrastructure in AI development and deployment. Third, AI model integration has also become an increasingly important focus, exemplified by HF's collaborations with AI companies like Meta AI and Stability AI, which aim to offer day-zero integrations for new model releases, capitalising on first-mover advantages and maintaining competitive edges in the rapidly evolving AI industry.

We acknowledge that our quantitative findings are temporally limited to September 2022, predating major industry developments like the launch of ChatGPT \cite{openai_introducing_2022}, the release of open-weight foundation models like Meta's LLaMA models \cite{touvron_llama_2023}, and grassroots initiatives like the BigScience Workshop \cite{akiki_bigscience_2022}. Given the pace of change in the AI industry, it is likely that the population of contributing companies and the collaboration structures within the projects have evolved since 2022. However, the nature and extent of these changes may vary across different layers of the AI stack. For example, at the deep learning framework level, since Meta's donation of PyTorch to the LF, the project has been governed by a governing board comprising representatives from AMD, Amazon Web Services, Google Cloud, HF, IBM, Intel, Meta, Microsoft, and Nvidia \cite{osborne_why_2024}. This governance change has not resulted in significant net increases in contributions, but it has seen an increase in contributions by AI accelerator manufacturers and a decrease by Meta \cite{yue_igniting_2024}.  

By contrast, Transformers has likely seen more significant developments to its community of contributors. The growth of the HF Hub as the \textit{de facto} platform for sharing and hosting open-weight models, has likely influenced participation in the development of Transformers, which provides APIs to download, use, and share models on the HF Hub. For example, the interviews revealed that since data collection, HF has engaged in strategic collaborations with AI model providers like Meta AI and Stability as well as cloud compute providers like Microsoft Azure, AWS, and Google Cloud, who have become active contributors to the project. Furthermore, Transformers' integration with multiple deep learning libraries (Jax, PyTorch, and TensorFlow) and its role in facilitating access to open-weight models hosted on the HF Hub likely position it at the centre of evolving collaboration dynamics in the open source AI ecosystem.

\subsection{Threats to Validity}\label{sec:coop-threatstovalidity}
\subsubsection{Construct validity}
Construct validity concerns the extent to which the measurements accurately represent the phenomenon under study \cite{easterbrook_selecting_2008}. We faced several challenges in this regard. First, while we operationalised commits to common files per release as a proxy for collaboration, as per prior work  \cite{zhang_how_2020,snarby_collaboration_2013},  we acknowledge this captures only one type of observable collaboration in OSS repositories among other types of collaboration \cite{casari_beyond_2023}. Moreover, by aggregating network data, we sacrificed granularity to enable a more manageable year-to-year comparative analysis of inter-company collaboration. Second, identifying the affiliations of individual contributors was challenging. Domain-name mining from email addresses, considered the most accurate method \cite{nguyen-duc_software_2019,mehra_firms_2011}, identified affiliations for only 37.0\%, 48.9\%, and 9.2\% of contributors in PyTorch, TensorFlow, and Transformers respectively. These low numbers may be due to the use of Github privacy features and personal email addresses, as well as the prevalence of contributors who are indeed volunteers (especially in Transformers). Manual labelling was resource-intensive and may have introduced errors (e.g., by assigning a company affiliation to a contributor rather than volunteer status). Third, the SNA may have been influenced by the different network sizes. To account for such effects, we examined size-independent network metrics across the projects and over time (see Tables~\ref{tab:network-metrics-pytorch}-\ref{tab:network-metrics-transformers}), which corroborated the SNA findings. Fourth, recruitment challenges limited our interview sample ($n=10$, only $n=2$ respondents for Transformers). Nonetheless, the interviews provided insights from various seniority levels and company profiles, and positive feedback on the quantitative findings during the interviews increased our confidence in the validity of the findings.

\subsubsection{External validity}

External validity concerns the generalisability of the findings. We acknowledge three key threats. First, the case study research design limits the generalisability of the findings. However, we underscore that the objective was to test theory, and the case study research design was suitable for this objective \cite{yin_case_2018}. Furthermore, the inclusion of Transformers expanded the analysis beyond industry giants' projects, addressing a limitation of prior work \cite{orucevic-alagic_network_2014, teixeira_collaboration_2014}. Second, the findings are based on a limited sample of projects ($n=3$) and interviews ($n=10$) and may not be exhaustive or universally applicable across all company-hosted OSS projects. Further research is needed to validate these findings across a wider range of projects, companies, and sectors. Furthermore, the interview sample does not capture the diversity of views of the broader population of contributors or senior decision-makers. Nonetheless, the sample included various seniority levels, from junior developers to start-up founders, as well as various company profiles, from start-ups to industry giants, thus providing a diversity of perspectives. Third, the quantitative findings were temporally limited to the data collection cut-off in September 2022, predating developments like the launch of ChatGPT \cite{openai_introducing_2022}, open-weight models like LlaMa 2 \cite{touvron_llama_2023}, or the rising popularity of the HF Hub \cite{osborne_ai_2024}, among others. Given the pace of change in the AI industry, we must assume that the projects have evolved since data collection. However, since our research objective was to test prior theory on open source co-opetition, this limitation is defensible. While some findings are specific to the AI industry, the identified collaborative relationship types, practices, and governance challenges likely remain relevant despite recent technological advancements in AI and beyond.

\subsubsection{Reliability}
Challenges to the reliability concern to the extent of the study's repeatability. With regards to the quantitative analysis, the Python scripts and Jupyter notebooks used for data collection and analysis are available in a public GitHub repository \cite{osborne_python_2024}. The involvement of three authors in the manual labelling and validation of contributors' company affiliations was a labour-intensive process that took one working week. While this procedure improved data quality, the labour intensity decreases the repeatability of this procedure for both the examined case studies as well as new cases in the future. With regards to the interviews, the interviews were recorded and transcribed by the first author to aid the analysis, and research guidelines were followed for the thematic analysis of the interview transcripts \cite{braun_using_2006}. Moreover, the involvement of two authors in coding and validation of codes reduced the potential biases that may arise when a single author performs qualitative data analysis alone \cite{cruzes_recommended_2011}. Finally, we member-checked findings with respondents to increase the reliability and practical relevance of the findings \cite{lincoln_naturalistic_1985}.

\section{Conclusion} \label{sec:coop-conclusion}

This study investigated how companies collaborate in company-hosted OSS projects via a mixed-methods analysis of Meta's PyTorch prior to its donation to the LF, Google's TensorFlow, and HF's Transformers. Overall, we make three key contributions that extend the literature on open source co-opetition. First, while the projects exhibit similar code authorship patterns between host and external companies (\textasciitilde80/20\% of commits respectively), collaborations are structured differently (e.g., decentralised vs. hub-and-spoke networks). Second, host and external companies engage in strategic, non-strategic, and contractual collaborations, which vary in the relevance of business strategy, competitive dynamics, and personal incentives of the involved developers. Some of the observed collaborations are specific to the AI industry (e.g., hardware-software optimizations or AI model integrations), while others are typical of the broader software industry (e.g., bug fixing or task outsourcing). Third, single vendor governance in company-hosted OSS projects creates a power imbalance that shapes open source co-opetition practices and possibilities, from a host company's singular decision-making power (e.g., risk of license change) to its community involvement strategy (e.g., from over-control to over-delegation). We concluded with recommendations for future research to advance our understanding of commercial participation and its impact on collaboration dynamics in OSS communities.

\newpage

\bibliographystyle{ieeetr}  
\bibliography{references}

\section{Acknowledgements}
Cailean Osborne was funded by the UK Economic and Social Research Council's Grand Union Doctoral Training Partnership Digital Social Science Pathway (Grant Number: ES/P000649/1). The remaining authors were funded by National Natural Science Foundation of China (Grant Number: 62332001). We would like to thank Johan Linåker, Vili Lehdonvirta, Fabian Stephany, Mark Graham, Xiaowen Dong, and the anonymous ACM CSCW reviewers for their feedback on prior versions of the manuscript.

\newpage
\appendix
\section{Commercial dominance in PyTorch, TensorFlow, and Transformers}

\begin{table}[ht]
	\centering
	\small
	\caption{Metrics for host company dominance in PyTorch, TensorFlow, and Transformers}
	\begin{tabular}{lccc} \toprule
     	\textbf{Metrics} & \textbf{PyTorch} & \textbf{TensorFlow} & \textbf{Transformers} \\ \midrule
    	\# developers & 738 & 1,172 & 174 \\
         \# host company employees & 452 (61.25\%) & 558 (47.61\%) & 56 (32.18\%) \\
         \# companies & 66 & 167 & 75 \\
    	\# developers that made 80\% of commits & 157 (12.54\%) & 248 (12.03\%) & 21 (7.54\%) \\
    	\# companies that made 80\% of commits & 1 (1.52\%) & 1 (0.59\%) & 1 (1.33\%) \\
    	\# bus factor (50\% commits) & 43 & 78 & 5 \\
    	\# bus factor (50\% net LOC) & 22 & 34 & 4 \\ \bottomrule
	\end{tabular}
	\label{tab:coop-project-dominance}
	\footnotesize
 
    \textit{N.B.: Limited to contributors who have made $\geq$ 5 commits.  The bus factor \\ is the smallest number of people that make 50\% of contributions \cite{chaoss_metric_2023}.}
\end{table}

\newpage
\section{Top Corporate Contributors to the OSS projects}
\label{sec:coop-top10companies}
\begin{table}[htbp]
\begin{minipage}{\textwidth}
\centering
\small
\caption{Top ten corporate contributors to PyTorch ranked by \texttt{n\_commits}}
\begin{tabular}{lcclcc} \toprule
\texttt{Affiliation} & \texttt{Commits} & \texttt{LOC (net)}&  \texttt{Out-degree} &\texttt{PageRank}  & \texttt{Betweenness} \\
\midrule
Meta & 0.84 & 0.84 &  0.98 &0.39  & 0.02 \\
Quansight & 0.05 & 0.04 &  0.67 &0.14  & 0.25 \\
Nvidia & 0.03 & 0.04 &  0.78 &0.11  & 0.31 \\
Microsoft & 0.02 & 0.04 &  0.51 &0.02  & 0.11 \\
AMD & 0.01 & 0.01 &  0.59 &0.04  & 0.16 \\
Google & 0.01 & 0.01 &  0.73 &0.05  & 0.20 \\
Intel & 0.01 & 0.01 &  0.65 &0.04  & 0.35 \\
IBM & 0.01 & <0.01 &  0.35 &0.01  & 0.04 \\
Twitter & <0.01 & <0.01 &  0.33 &0.01  & <0.01 \\
DeepMind & <0.01 & <0.01 &  0.16 &0.01  & <0.01 \\
\bottomrule
\end{tabular}
\label{tab:coop-pytorch-top-companies}
\footnotesize 

\textit{N.B. Contributions by volunteers (\texttt{Commits}=0.05, \texttt{LOC}=0.05)  and \\ unknown affiliations (\texttt{Commits}=0.02, \texttt{LOC}=0.01) are excluded.}
\end{minipage}

\vspace{1em}

\begin{minipage}{\textwidth}
\centering
\small
\small
\caption{Top ten corporate contributors to TensorFlow ranked by \texttt{n\_commits}}
\begin{tabular}{lcclcc} \toprule
\texttt{Affiliation} & \texttt{Commits} & \texttt{LOC (net)} &  \texttt{Out-degree} &\texttt{PageRank}  & \texttt{Betweenness} \\
\midrule
Google & 0.85 & 0.34 &  0.97 &0.25  & 0.04 \\
Nvidia & 0.03 & 0.15 &  0.64 &0.04  & 0.07 \\
Intel & 0.03 & 0.12 &  0.86 &0.17  & 0.09 \\
IBM & 0.02 & 0.09 &  0.72 &0.05  & 0.06 \\
AMD & 0.01 & 0.06 &  0.49 &0.01  & 0.04 \\
Arm & 0.01 & 0.03 &  0.64 &0.10  & 0.04 \\
Huawei & 0.01 & 0.02 &  0.33 &0.01  & 0.07 \\
Microsoft & <0.01 & 0.02 &  0.39 &0.01  & 0.15 \\
Graphcore & <0.01 & 0.02 &  0.41 &0.01  & 0.12 \\
Offscale & <0.01 & 0.01 &  0.14 &<0.01  & 0.01 \\
\bottomrule
\end{tabular}
\label{tab:coop-tensorflow-top-companies} 
\footnotesize 

\textit{N.B. Contributions by volunteers (\texttt{Commits}=0.10, \texttt{LOC}=0.06)  and \\ unknown affiliations (\texttt{Commits}=0.04, \texttt{LOC}=0.11) are excluded.}
\end{minipage}

\vspace{1em}

\begin{minipage}{\textwidth}
\centering
\small
\caption{Top ten corporate contributors to Transformers ranked by \texttt{n\_commits}}
\begin{tabular}{lcclcc} \toprule
\texttt{Affiliation} & \texttt{Commits} & \texttt{LOC (net)} &  \texttt{Out-degree} &\texttt{PageRank}  & \texttt{Betweenness} \\
\midrule
HF & 0.91 & 0.94 &  0.96 &0.43  & 0.39 \\
Fractal Ideas & 0.01 & 0.01 &  0.17 &0.14  & 0.08 \\
Intel & 0.01 & 0.01 &  0.17 &0.01  & 0.08 \\
Saama AI Research & <0.01 & <0.01 &  0.13 &0.01  & 0.01 \\
Uber & <0.01 & <0.01 &  0.02 &<0.01  & <0.01 \\
Deepset & <0.01 & <0.01 &  0.09 &0.01  & 0.19 \\
Telekom & <0.01 & <0.01 &  0.06 &<0.01  & <0.01 \\
Oracle & <0.01 & <0.01 &  0.15 &0.01  & <0.01 \\
LTIMindtree & <0.01 & <0.01 &  0.17 &0.01  & 0.25 \\ 
Google & <0.01 & <0.01 &  0.09 &0.01  & 0.07 \\
\bottomrule
\end{tabular}
\label{tab:coop-transformers-top-companies}
\footnotesize 

\textit{N.B. Contributions by volunteers (\texttt{Commits}=0.10, \texttt{LOC}=0.06)  and \\ unknown affiliations (\texttt{Commits}=0.01, \texttt{LOC}=0.01) are excluded.}
\end{minipage}
\end{table}

\newpage
\section{Network Metrics for Collaboration on PyTorch, TensorFlow, and Transformers}

\begin{table}[htbp]
\begin{minipage}{\textwidth}
\centering
\small
\caption{Network Metrics for PyTorch}
\begin{tabular}{lccccc} \toprule
\texttt{Year} & \texttt{Nodes} & \texttt{Edges} & \texttt{Deg Cent} & \texttt{Deg Skew} & \texttt{Clust Coef} \\
\midrule
2016 & 9 & 34 & 0.5938 & 0.16 & 0.5579 \\
2017 & 20 & 130 & 0.6925 & 0.64 & 0.6202 \\
2018 & 26 & 194 & 0.7296 & 1.44 & 0.7158 \\
2019 & 24 & 178 & 0.6616 & 0.89 & 0.7096 \\
2020 & 25 & 210 & 0.6337 & 0.65 & 0.7802 \\
2021 & 16 & 108 & 0.5156 & 0.23 & 0.7827 \\
2022 & 16 & 126 & 0.4356 & -0.26 & 0.6956 \\
\bottomrule
\end{tabular}
\label{tab:network-metrics-pytorch}
\end{minipage}
\vspace{1em}
\begin{minipage}{\textwidth}
\centering
\small
\caption{Network Metrics for TensorFlow}
\begin{tabular}{lccccc} \toprule
\texttt{Year} & \texttt{Nodes} & \texttt{Edges} & \texttt{Deg Centr} & \texttt{Deg Skew} & \texttt{Clust Coef} \\
\midrule
2016 & 22 & 104 & 0.8118 & 2.68 & 0.6843 \\
2017 & 43 & 276 & 0.8673 & 2.89 & 0.6836 \\
2018 & 43 & 370 & 0.7653 & 1.86 & 0.7801 \\
2019 & 55 & 810 & 0.7407 & 1.65 & 0.8160 \\
2020 & 63 & 1208 & 0.6855 & 1.33 & 0.7659 \\
2021 & 46 & 692 & 0.6123 & 0.94 & 0.7882 \\
2022 & 24 & 190 & 0.5028 & 0.78 & 0.7634 \\
\bottomrule
\end{tabular}
\label{tab:network-metrics-tensorflow}
\end{minipage}
\vspace{1em}
\begin{minipage}{\textwidth}
\centering
\small
\caption{Network Metrics for Transformers}
\begin{tabular}{lccccc} \toprule
\texttt{Year} & \texttt{Nodes} & \texttt{Edges} & \texttt{Deg Centr} & \texttt{Deg Skew} & \texttt{Clust Coef} \\
\midrule
2018 & 4 & 10 & 0.2222 & 0.00 & 0.8333 \\
2019 & 23 & 70 & 0.9008 & 3.81 & 0.5155 \\
2020 & 27 & 68 & 0.9379 & 4.62 & 0.3342 \\
2021 & 27 & 100 & 0.8506 & 3.13 & 0.4387 \\
2022 & 24 & 82 & 0.7977 & 3.39 & 0.4756 \\
\bottomrule
\end{tabular}
\label{tab:network-metrics-transformers}
\footnotesize 

\textit{N.B. Deg Cent = Degree Centralisation, Deg Skew = Degree Skew, Clust Coef = Clustering Coefficient. \\ Since the networks are directed networks with reciprocated edges,  we report a single degree centralisation value \\ rather than separate in- and out-degree centralisation values due to their identical values.}
\end{minipage}
\vspace{1em}
\end{table}

\end{document}